\documentclass[twocolumn,showpacs,preprintnumbers,amsmath,amssymb, prb, reprint,showkeys]{revtex4-1}
\usepackage{amsmath}
\usepackage{stmaryrd}
\usepackage{txfonts}
\usepackage{amssymb}
\usepackage{mathrsfs}
\usepackage{graphicx}
\usepackage{dcolumn}
\usepackage{bm}
\usepackage{epsfig}
\usepackage{color}
\usepackage[colorlinks=true, linkcolor=blue, citecolor=blue, urlcolor=blue]{hyperref} 
\usepackage{epstopdf}

\begin{document}

\title{Density functional theory study of skyrmion pinning by atomic defects in MnSi}

\author{Hong Chul Choi}
%%\email{ithink@postech.ac.kr}
\affiliation{Theoretical Division, Los Alamos National Laboratory, Los Alamos, New Mexico 87545, USA}
\author{Shi-Zeng Lin}
%%\email{ithink@postech.ac.kr}
\affiliation{Theoretical Division, Los Alamos National Laboratory, Los Alamos, New Mexico 87545, USA}
\author{Jian-Xin Zhu}
%%\email{ithink@postech.ac.kr}
\affiliation{Theoretical Division, Los Alamos National Laboratory, Los Alamos, New Mexico 87545, USA}
\affiliation{Center for Integrated Nanotechnologies, Los Alamos National Laboratory, Los Alamos, New Mexico 87545, USA}
\begin{abstract}
A magnetic skyrmion observed experimentally in chiral magnets is a topologically protected spin texture. For their unique properties, such as high mobility under current drive, skyrmions have huge potential for applications in next-generation spintronic devices. Defects naturally occurring in magnets have profound effects on the static and dynamical properties of skyrmions. In this work, we study the effect of an atomic defect on a skyrmion using the first-principles calculations within the density functional theory, taking MnSi as an example. By substituting one site of Mn or Si with different elements, we can tune the pinning energy. The effects of pinning by an atomic defect can be understood qualitatively within a phenomenological model. 
\end{abstract}
	
%12.39.Dc	Skyrmions
%73.20.Hb	Impurity and defect levels; energy states of absorbed species
%71.15.Nc Total energy and cohesive energy calculations
\date{\today}
\maketitle
\section{Introduction}
A skyrmions in magnets is a spin texture with a topological charge $N=\int d^2r\mathbf{n}\cdot(\partial_x\mathbf{n}\times\partial_y\mathbf{n})/(4\pi)=\pm 1$, where the unit vector $\mathbf{n}$ represents the direction of local magnetic moments. Skyrmions in chiral magnets without inversion symmetry were predicted in 1992~\cite{Bogdanov89} and were later discovered in MnSi by small angle neutron scattering.~\cite{Muhlbauer2009} Real space imaging by Lorentz transmission electron microscopy was performed in thin films and has revealed a detailed spin arrangement in an individual skyrmion~\cite{Yu2010a}. Furthermore a dilute gas of skyrmions was observed near phase boundary between the skyrmion lattice and ferromagnetic state, where the systems undergo a first-order phase transition between these two states. Since then skyrmions have been observed in many compounds, including magnetic 
metals,~\cite{Muhlbauer2009,Yu2010a} semiconductors~\cite{Yu2011} and insulators,~\cite{Adams2012,Seki2012} suggesting that skyrmions may be ubiquitous in magnets. In these magnets, the spatial inversion symmetry is broken, where the Dzyaloshinskii-Moriya (DM) 
interaction~\cite{Dzyaloshinsky1958,Moriya60,Moriya60b} is responsible for the stabilization of skyrmions. Because of the weakness of  the DM interaction compared to the ferromagnetic exchange interaction, the size of skyrmion is much bigger than the atomic lattice constant and is typically about tens of nanometers. 

Skyrmions can be manipulated by various external fields, such as electric and magnetic fields, temperature gradient and electric current 
etc.~\cite{Jonietz2010,Yu2012,Schulz2012,White2012,PhysRevLett.113.107203,Shibata2015} The ability to drive skyrmions by an electric current is particularly interesting from the viewpoint of spintronic applications. Remarkably the threshold current to drive the skyrmion is about $10^6\ \mathrm{A/m^2}$ which is about 5 to 6 orders of magnitude weaker than that for a magnetic domain wall. \cite{Jonietz2010,Yu2012,Schulz2012} There are several interesting theoretical proposal to utilize skyrmions in memory devices.~\cite{Fert2013,zhou_reversible_2014,sampaio_nucleation_2013} To this end, it is crucial to understand and control the dynamics of skyrmions. 

Defects occurring naturally in real materials provide a pinning barrier for skyrmions and is responsible for the experimental observed threshold current to drive skyrmions into motion. Similar to vortices in superconductors~\cite{Blatter94}, defects impact the behavior of skyrmions both in equilibrium and dynamically, such as arrangement of skyrmions and direction of motion of skyrmions. For applications, it is necessary to control the energy landscape generated by defects. For instance, one needs to pin skyrmions at a desired position in the memory applications. Therefore, it is required to understand origin of pinning of skyrmions by defects. There are several work, where the effects of defects were modeled 
phenomenologically,~\cite{szlin13skyrmion2,Everschor12,Iwasaki2013} but the microscopic mechanism of pinning effect has not been considered. In this work, we address this issue by performing the first-principles calculations within the density functional theory (DFT). We study the effect of an atomic defect on a single skyrmion by focusing on the prototypical skyrmion-hosting materials MnSi as an example. We show that the defect energy landscape can be tuned by substitution of Mn or Si with different elements. The impurity modifies the local electronic density of states (LDOS) and spin-orbit coupling, which in turn change the interactions between magnetic moments. We then provide a qualitative understanding of the pinning effect based on a phenomenological model.

\section{Method}

\begin{figure}[tb]
\includegraphics[width=1.0\linewidth]{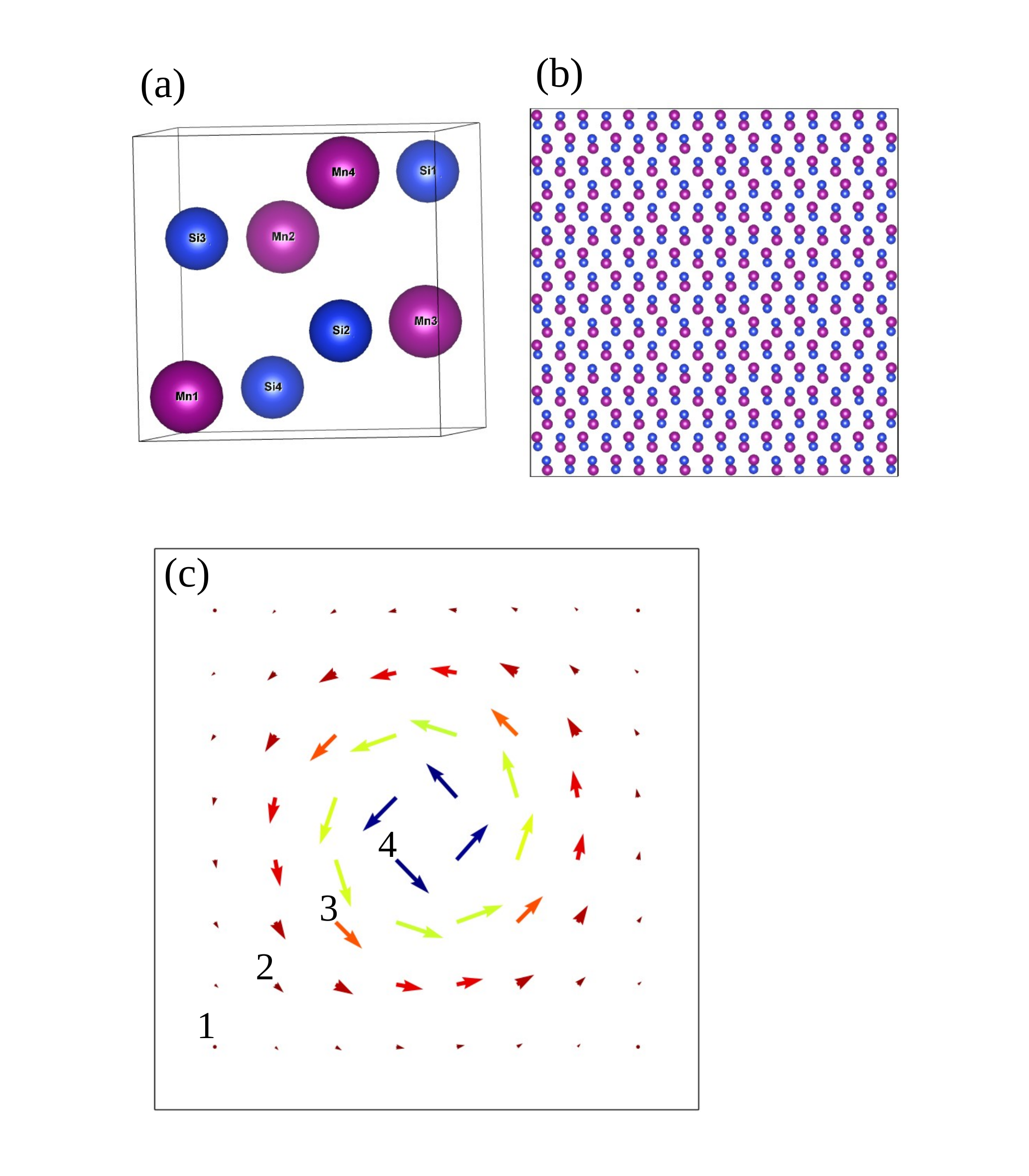}
\caption{(color online)
%A primitive unitcell of MnSi and the supercell used for the calculation.
(a) Primitive unit cell of MnSi has 4 Mn and 4 Si atoms in the cubic symmetry. (b) The $8\times8\times1$ supercell was adopted to accommodate
 a skyrmion spin texture with 256 Mn atoms and 256 Si atoms. (c) The four Mn atoms in a unit cell are located at $z=$0.1395, 0.3605, 0.6395, and 0.8605, respectively. The $8\times8\times1$ supercell can be regarded as four  $8\times8$ two dimensional (2D) lattices stacked along the $z$-axis. We label the position at $(1, 1)$, $(2, 2)$, $(3, 3)$, and $(4,4)$ in the 2D coordinate as 1, 2, 3, 4  respectively, for convenience.
}
\label{fig1}
\end{figure}

The cubic primitive cell (lattice constant= 0.4559 nm) of  $\mathrm{MnSi}$, which belongs to the $P{2}_{1} 3$ space group symmetry, is displayed in Fig. \ref{fig1}(a).  The primitive cell has 4 $\mathrm{Mn}$ and 4 $\mathrm{Si}$ atoms without an inversion center. The lack of an inversion symmetry together with the spin-orbit coupling gives rise to  the DM interaction. There is also ferromagnetic exchange interaction between Mn atoms due to the double-exchange mechanism mediated by conduction electrons. The competition between the weak DM interaction and exchange interaction stabilizes a long wavelength magnetic normal spiral in the ground state, with a pitch period $\lambda=$18 nm in bulk and 8.5 nm in thin films. In bulk, the normal spiral becomes the conical spiral state under moderate magnetic field of the order of $0.1$ Tesla in the most region of the magnetic field-temperature phase diagram. Near $T_N=30$ K, a triangular lattice of skyrmion is stabilized in a small region of the phase diagram. In thin films, the skyrmion phase is stabilized down to zero temperature.

We perform first-principles calculations within the DFT to study the effect of an impurity on a skyrmion in MnSi, employing the projector augmented plane-wave method \cite{paw} implemented in Vienna {\it ab-initio} simulation package (VASP)~\cite{vasp}. 
We used the VASP pseudopotential (Mn, Si), a $1\times1\times1$ Monkhorst-pack-kpoint mesh, 
and a 300 eV cutoff.
The linear mixing method for updating spin-polarized electron density was adopted.
The spin-orbit coupling is accounted for to include the DM interaction in the calculation.

The skyrmion size in MnSi is about 10 nm. This requires a large number of unit cell, which renders the calculations extremely expensive. To make the problem tractable, we introduce a smaller system size with the periodic boundary condition and confine a skyrmion at the center of the unit cell. In this case, the size of skyrmion is determined by the competition of exchange, DM interactions and the geometry confinement due to the small system size.  In the calculations, we use $8\times 8\times 1$ supercell, as shown in Fig. \ref{fig1}(b). We then introduce a skyrmion into the system by patternizing the spin configuration on Mn atoms, $\mathbf{S}(x, y, z)$. A skyrmion is centrosymmetric and we use the polar coordinate $\mathbf{r}=(r, \phi)$. At its center, the spins point down while spins point up in the region away from the skyrmion center. The spins rotate in the azimuthal direction with the rotation direction fixed by the DM vector. We initialize the system according to
\begin{equation}\label{eqSxSySz}
(S_x, S_y, S_z)=S_0[\cos(\varphi) \sin(\theta),\ \sin(\varphi) \sin (\theta),\ \cos (\theta)],
\end{equation}
\begin{equation}\label{eqtheta}
\theta(r) = \pi [1 -{\tanh(r/r_{0})}],\ \ \ \mathrm{and}\ \ \  \varphi=\phi+\pi/2 ,
\end{equation}
In the above notation, we have projected all Mn into the $x$-$y$ plane. The phase shift $\pi/2$ accounts for the helicity of skyrmion in MnSi. The Si atoms do not carry magnetic moment actively. We then relax the system in the calculations until convergence is achieved. The system may be trapped by a local energy minimum. To obtain a lower energy state, we use different initial condition by changing $r_0$. The calculated magnetic moment of Mn does not depend on the initial value $S_0$.

The definition of skyrmion topological charge $N=\int d^2r\mathbf{n}\cdot(\partial_x\mathbf{n}\times\partial_y\mathbf{n})/(4\pi)$ valid in the continuum limit does not apply here because of the small skyrmion size. We use an alternative definition by calculating the solid angle 
\begin{equation}
\Theta_i = 2\arctan\left[\frac{\mathbf{S}_{i} \cdot (\mathbf{S}_{j} \times \mathbf{ S}_{k})} {1+\mathbf{S}_i\cdot\mathbf{S}_j +\mathbf{S}_j\cdot\mathbf{S}_k +\mathbf{S}_k\cdot\mathbf{S}_i} \right],
\end{equation}
subtended by three neighboring spins, $\mathbf{S}_i$ in the projects $x$-$y$ plane. We then sum $\Theta_i$ in the whole system and the skyrmion topological charge is given by $N_S=\sum_i\Theta_i/4\pi$. Here $N_s=1$ for the skyrmion considered in Eqs.~\eqref{eqSxSySz} and~\eqref{eqtheta}.

The calculated magnetic moment of a Mn ion is 1 $\mu_{B}$, which is larger than the experimental value 0.4 $\mu_{B}$,~\cite{Ishikawa1976} which is consistent with the previous first-principles calculations for a ferromagnetic state of MnSi.~\cite{MnSi-DFT-04} The overestimate of the magnetic moment may be due to the strong correlation effect and quantum fluctuation of moments neglected in the DFT calculations. Despite the overestimate of moment and the finite size effect, we hope that the qualitative features of the effect of an atomic impurity on a skyrmion can be captured by the present study and provides guidance to understand microscopic origin of the pinning of skyrmions.

In the calculations, we fix the center of a skyrmion at the center of the supercell. We substitute one Mn or Si at different site by Co, Ir, Zn and Pb and calculate the total energy. No external magnetic field is applied in the calculations.

\section{Computational results}

\begin{figure}[tb]
\includegraphics[width=1.0\linewidth]{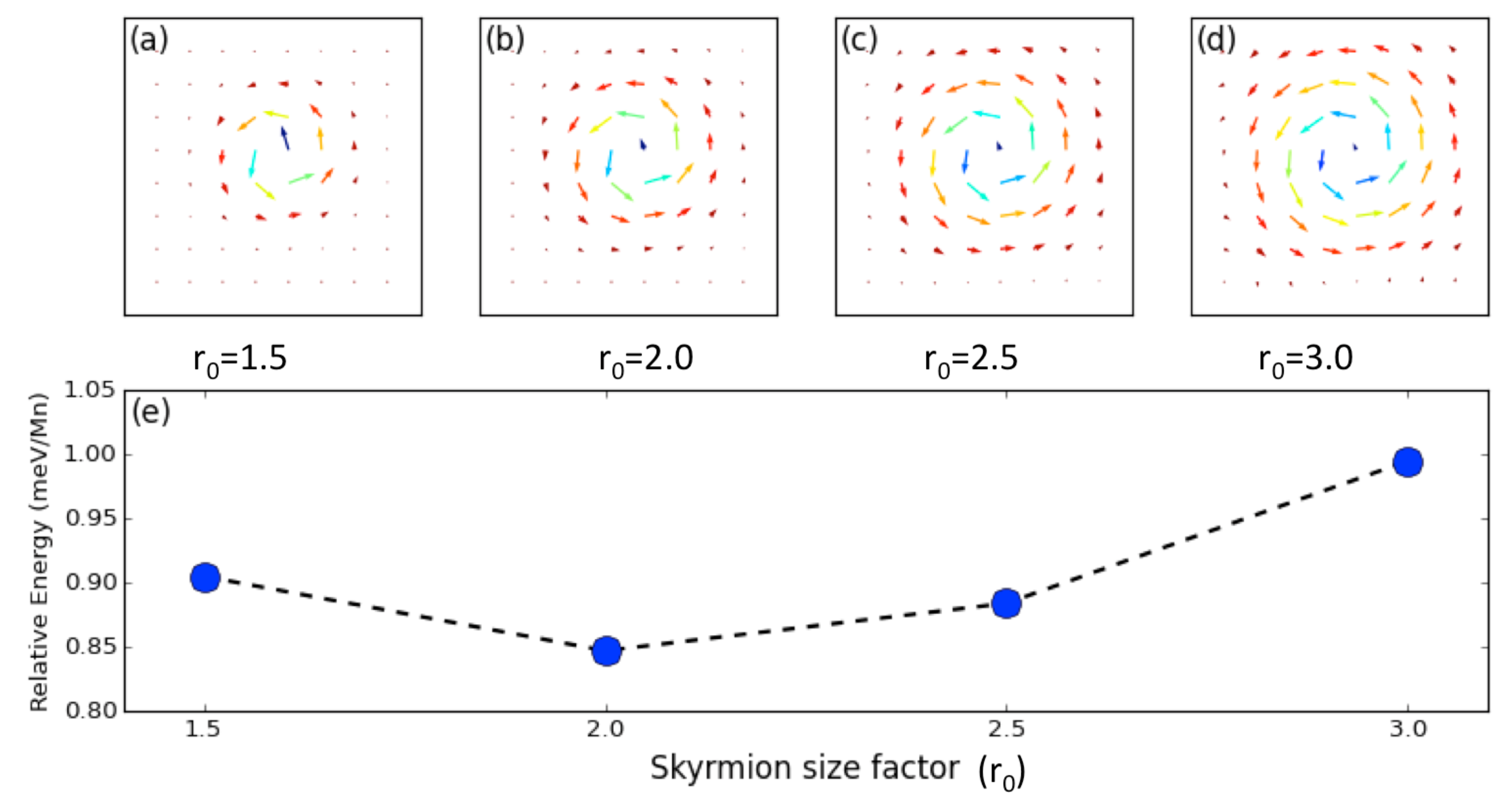}
\caption{(color online)
(a)-(d) Skyrmion spin textures and (e) total energy of a skyrmion relative to the energy in the ferromagnetic state as a function of the initial skyrmion size $r_0$ in Eq.~\eqref{eqtheta} in the $8\times8\times1$ supercell.
In Fig. (a)-(d), arrows demonstrate the in-plane component of local moments at each site.
The blue (red) color of the arrow  means a negative (positive) value of the $z$ component of the local moment.
}
\label{fig2}
\end{figure}

Before proceeding to investigate the effect of an impurity, we first obtain the optimal skyrmion texture in the $8\times8\times1$ supercell of MnSi by starting from different initial conditions parameterized by $r_0$. The dependence of the total energy and the corresponding skyrmion structure are shown in 
Fig.~\ref{fig2}. The optimal skyrmion size in this supercell is about $r_0=2$. This can be understood as follows. For a small skyrmion size, such as $r_0=1.5$, it costs energy in the ferromagnetic exchange interaction because the skyrmion deviates significantly from its optimal size determined by the competition between the exchange and DM interaction, which is much bigger than the supercell size. On the other hand, when skyrmions size increases such as those shown in Figs. \ref{fig2}(c) and (d),  the spins are not parallel and have in-plane components that wind counterclockwise at the boundary. This costs energy because the in-plane components of spins are antiparallel at the boundary due to the periodic arrangement of the supercell. The skyrmion energy is higher than that of a ferromagnetic state.

\begin{figure}[b]
\includegraphics[width=1.0\linewidth]{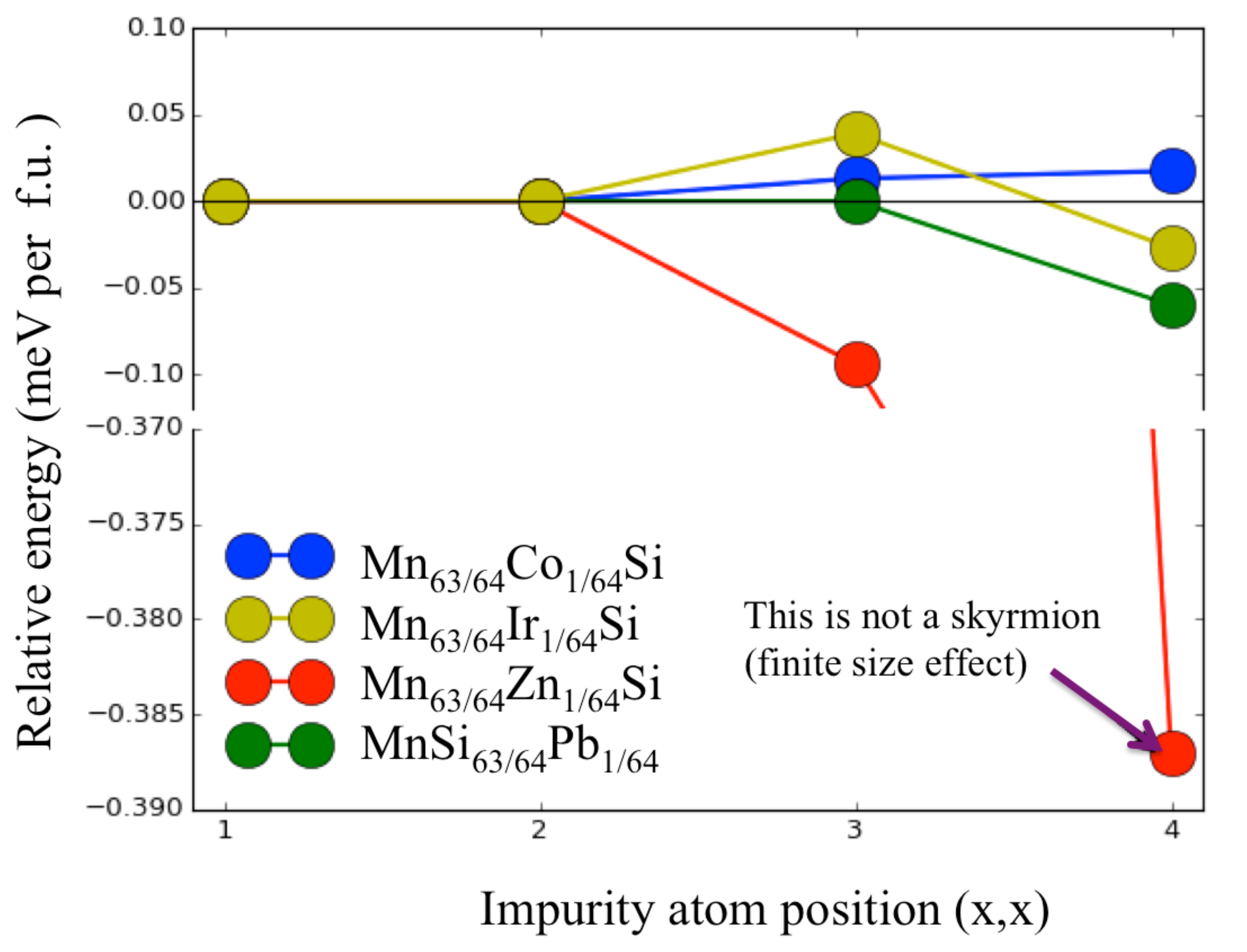}
\caption{(color online) Relative total energy as function of the impurity  position for various impurity atoms in the $8\times8\times1$ supercell. Here we take the skyrmion energy with an impurity atom located at (1,1) as an energy reference. All the spin configurations have the skyrmion topological charge $N_s=1$ except for $\mathrm{Mn_{63/64}Zn_{1/64}Si}$ with $N_s\approx 0.5$. The 2D coordinate is defined Fig. \ref{fig1}(c).
}
\label{fig3}
\end{figure}

After obtaining a metastable skyrmion solution by the DFT calculations, with all the microscopic interactions being taken into account, we then introduce an atomic impurity by replacing one of the Mn or Si atom in a supercell by an alien atom. Because of the centrosymmetry of the skyrmion, we introduce an impurity at a varying distance from the skyrmion center, as labeled by 1,2, 3, and 4 in Fig \ref{fig1}(c). When one of 64 formula unit atoms is substituted with one impurity atom, such as Co, Ir, Zn and Pb considered here, the doping concentration becomes 1/64, which corresponds to $\mathrm{Mn_{63/64}Co_{1/64}Si}$, $\mathrm{Mn_{63/64}Ir_{1/64}Si}$, $\mathrm{Mn_{63/64}Zn_{1/64}Si}$, and $\mathrm{MnSi_{63/64}Pb_{1/64}}$. The total energy as a function impurity position for different impurity atom is depicted in Fig. \ref{fig3}. For the Zn and Pb impurities, the energy decreases when the impurity is close to the skyrmion center, indicating an attraction between the skyrmion and the impurity. Therefore the Zn and Pb impurities behave as pinning centers for skyrmions. On the other hand, for the Co impurity, the energy increases meaning a repulsive interaction between the skyrmion and the impurity. The Co impurity works as an energy barrier for skyrmions. For the Ir impurity, the interaction between the impurity and the skyrmion is nonmonotonic. When the impurity approaches the skyrmion center, the interaction is repulsive and then becomes attractive when the impurity is at the skyrmion center. A qualitative understanding of these behavior will be presented below.

The Zn doped at the skyrmion center has dramatic effect as indicated by a sharp decrease of energy in Fig. \ref{fig3}. The calculated skyrmion topological charge is $N_s\approx 0.5$. In real systems with a large skyrmion size, a single atomic impurity cannot modify the whole skyrmion texture because a skyrmion is topologically protected against local perturbations. Therefore we ascribe the reduction of $N_s$ here to the finite size effect. In real systems, the skyrmion energy is expected to decreases monotonically when the skyrmion approaches to the Zn impurity.

\begin{figure}[tb]
\includegraphics[width=1.0\linewidth]{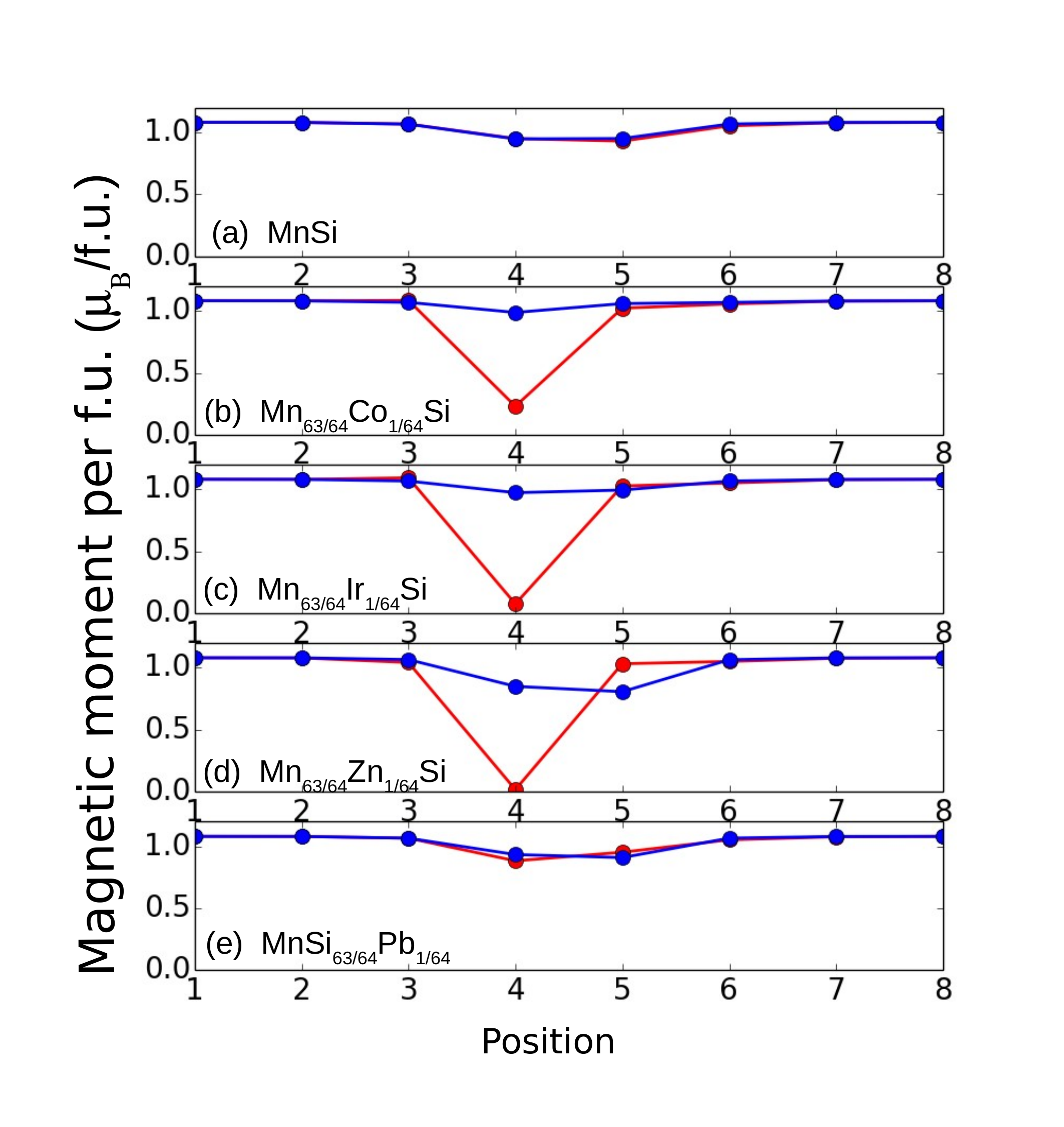}
\caption{(color online)
Magnetic moment $S$ as a function of the position in (a) $\mathrm{MnSi}$ and doped $\mathrm{MnSi}$ such as (b) $\mathrm{Mn_{63/64}Co_{1/64}Si}$, (c) $\mathrm{Mn_{63/64}Ir_{1/64}Si}$, (d) $\mathrm{Mn_{63/64}Zn_{1/64}Si}$, and (e) $\mathrm{MnSi_{63/64}Pb_{1/64}}$. The impurity atoms are located at (4,4), which are indicated by the small arrow.  The red and blue lines represent the different diagonals $(x,\ x)$ and $(x,\ 9-x)$, as ${x}$ changes from 1 to 8.
}
\label{fig4}
\end{figure}

The spatial variations of the local moments $S$ are investigated in MnSi and doped MnSi. 
The red and blue lines in Fig. \ref{fig4} show how the $S$ of each local moment  is distributed along the two different diagonals when the impurity atoms (Co, Ir, Zn, Pb) are placed at (4,4).
The magnetic moment of the localized $d$ electrons seems to be hardly affected by the $\mathrm{Pb}$ atom.
The magnetic moment is mainly determined by on-site interactions, and the screening effect of $p$ electrons has weak impact on the moments.
As Co and Ir atoms have two more valence electrons compared to Mn atoms,
the magnetic moments with more  than half filled occupancy 
at the impurity sites are reduced.
Zn atom has the fully-occupied $d$ orbital and it can also influence 
the reduction of the local moments of the neighboring sites. 
The local moment at (4,3) is also suppressed than that in $\mathrm{MnSi}$ , 
as shown in Fig.~\ref{fig4}(d).

\begin{figure}[tb]
\includegraphics[width=1.0\linewidth]{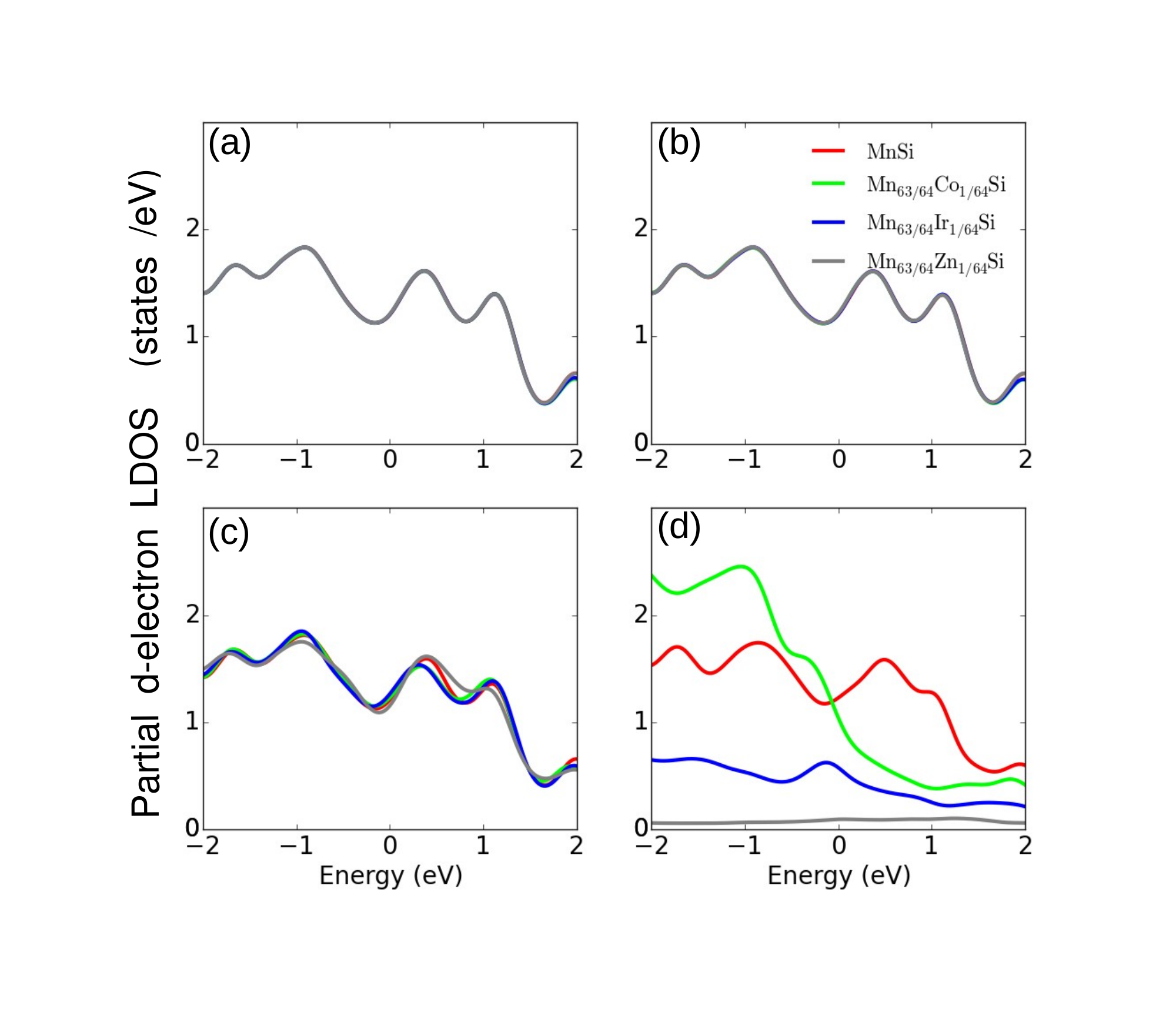}
\caption{(color online)
LDOS at different sites: (a) the boundary (1, 1), (b) (2, 2), (c)  (3, 3), and (d) the center (4, 4). The impurity is located at the skyrmion center (4, 4). For comparison, the LDOS of the pure MnSi system is also shown. The 2D coordinate for the supercell is defined in Fig.~\ref{fig1}(c).}
\label{fig5}
\end{figure}

The LDOS can provide information about the real-space electronic structure, which can be measured in experiments, such as the scanning tunneling microscopy. Meanwhile the interaction between magnetic moments depends on the LDOS, the knowledge of LDOS is helpful to understand the effect of impurity on skyrmion. Here we consider the effect of skyrmion spin texture on the LDOS in MnSi with an impurity.

Figure~\ref{fig5} demonstrates the LDOS at (1, 1)-(4, 4) from the boundary to center
for the various doped $\mathrm{MnSi}$'s, where the impurity atom is located at  (4, 4).
The LDOS at (1, 1) and (2, 2) are almost identical, 
because they are away from the impurity site. The LDOS at (3, 3) shows the slight difference,
as shown in Fig.~\ref{fig5}(c). The impurity has strong effect on the LDOS at the impurity site as displayed in Fig.~\ref{fig5}(d). The LDOS at the Fermi energy $E_F$ is reduced for the Co, Ir and Zn impurities in comparison to the case of the pure MnSi system. This behavior can be understood as follows. The energy level of the fully occupied $d$ electrons in the Zn ion is located below $E_F$ so that the $d$ states of Zn hardly contribute to $E_F$. For Ir, because $5d$ electrons are much more extended, which leads to a broader effective bandwidth and a reduced intensity of LDOS at the Fermi energy; while for Co, the $3d$ electrons are more localized than those of Ir $5d$ and close to Mn $3d$ electrons, therefore the intensity of LDOS on Co is similar to that on Mn.

\begin{figure}[tb]
\includegraphics[width=\columnwidth]{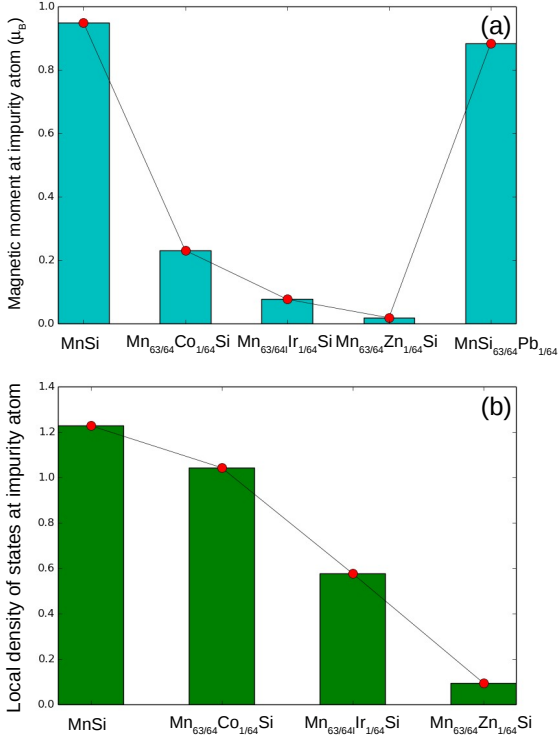}
\caption{(color online)
(a) Magnetic moment (b) LDOS at the impurity atom 
for $\mathrm{MnSi}$ and doped $\mathrm{MnSi}$, where the impurity atoms are placed at (4,4) in the $8\times8\times1$ supercell. The red circle and black lines are the peak positions and are guide to eyes.
}
\label{fig6}
\end{figure}

\begin{figure}[tb]
\includegraphics[width=1.0\linewidth]{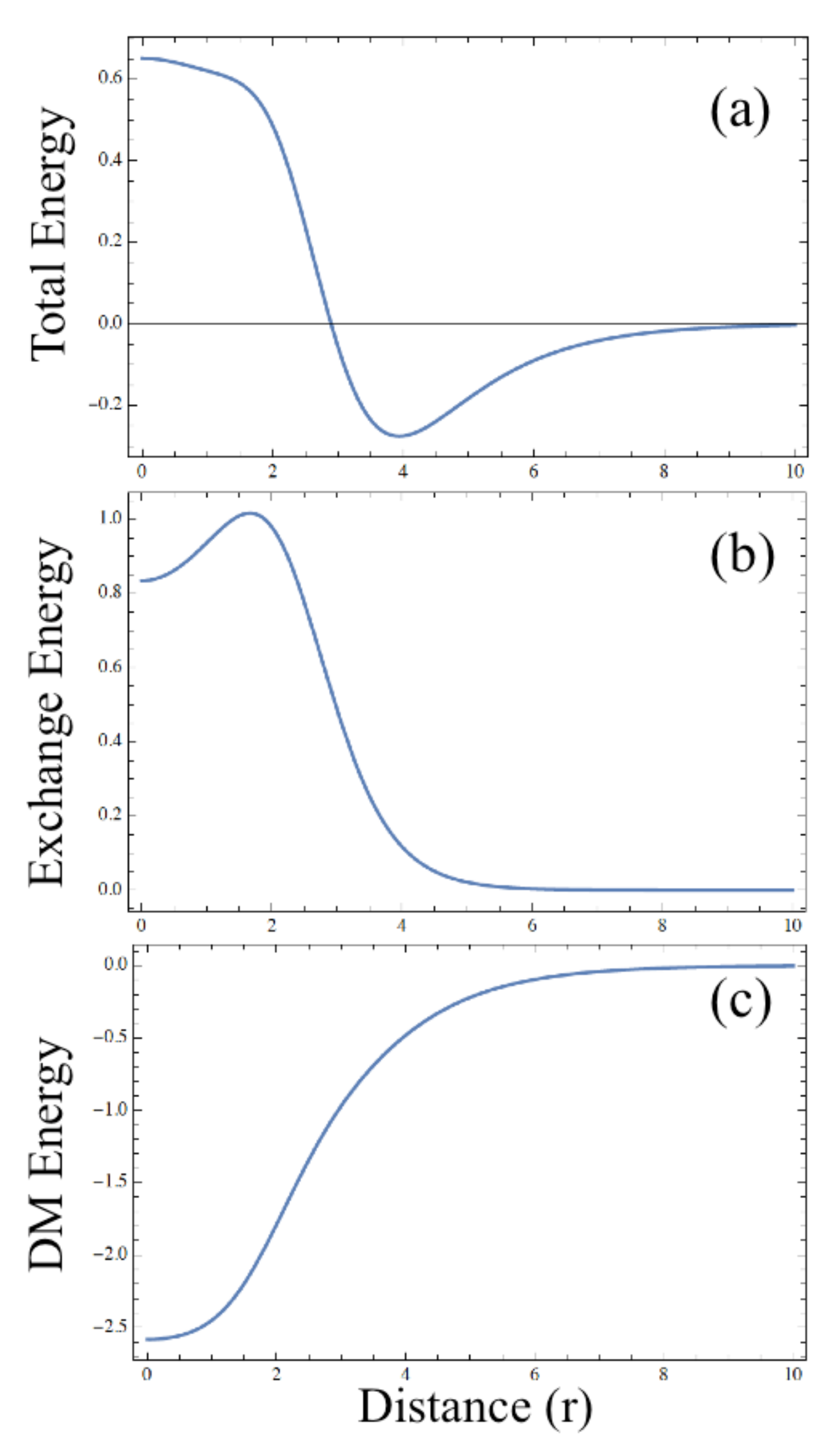}
\caption{(color online) (a) Total skyrmion energy relative to the ferromagnetic state, (b) exchange, and (c) DM interaction energies as a function of the distance from the center of the single skyrmion in pure systems. The energy is in unit of $D^2/J$ and the length is unit of $J/D$.}
\label{fig7}
\end{figure}

Figure~\ref{fig6} summarizes the local magnetic moment and LDOS  at Fermi surface on the impurity size with different atomic impurities in comparison to the pure system. The impurity is fixed at (4, 4) and the local moment is averaged over the four neighboring sites such as (4, 4), (4, 5), (5, 5),and (5, 4). The moment is suppressed in $\mathrm{Co}$, $\mathrm{Ir}$, and $\mathrm{Zn}$ doped $\mathrm{MnSi}$, while it does not change much in the  $\mathrm{Pb}$ doped $\mathrm{MnSi}$. The LDOS at $E_F$ in $\mathrm{Ir}$ and $\mathrm{Zn}$ doped $\mathrm{MnSi}$ are reduced less than half of that of $\mathrm{MnSi}$.
For the $\mathrm{Co}$ doped case, the reduction is about 10 $\%$ and is relatively small.

\section{Discussions}
Based on the DFT results, we now discuss the effect of impurities on the skyrmion using a phenomenological theory with an effective interaction between magnetic moments. The effective Hamiltonian density is
\begin{equation}\label{eqH}
\mathcal{H}= \frac{J}{2} \sum_{\mu=x, y} (\partial_{\mu} \mathbf{n})^2+D \mathbf{n} \cdot \nabla \times \mathbf{n} -\mathbf{B} \cdot \mathbf{n},
\end{equation}
where $J$ and $D$ are the exchange and DM interaction respectively. Equation~\eqref{eqH} has successfully captured several key experimental observations in B20 compounds.~\cite{okamura_microwave_2013,schwarze_universal_2015} At zero temperature, the Hamiltonian Eq.~\eqref{eqH} stabilizes a magnetic spiral, triangular lattice skyrmion and spin-polarized ferromagnetic state upon increasing the field. Both transitions are of the first order.~\cite{Bogdanov94}

Here we consider a single skyrmion solution in the ferromagnetic background in clean systems. The total energy and the contributions due to the exchange interaction, DM interaction as a function of distance $r$ from the skyrmion center are displayed in Fig.~\ref{fig7}. The single skyrmion is a metastable which has higher energy than the ferromagnet. The main energy contribution comes from the core region of the skyrmion. The skyrmion solution costs energy in the exchange interaction and save energy in the DM interaction. Based on this observation, an impurity can attract or pin a skyrmion if the exchange interaction or magnetic moment is reduced, or the DM interaction is enhanced. This observation provides a consistent understanding of the effects of an atomic impurity on the skyrmion obtained by the DFT calculations.   

For the Zn impurity, it reduces the magnetic moment and LDOS. For the Pb impurity, the magnetic moment and LDOS do not change much. Due to the large atomic number of Pb, the spin-orbit interaction, hence the DM interaction is enhanced. Therefore both the Zn impurity and Pb impurity provide attractive potential to the skyrmion. For the Ir defect, it also attracts skyrmion because the reduction of LDOS, magnetic moment and enhancement of DM interaction when it is doped at the skyrmion center. For the Co impurity, both the exchange and DM interactions are expected to be slightly reduced because of the reduction of the LDOS and magnetic moment. It could be possible that the energy loss due to the decrease of DM interactions outweigh the energy gain due to the reduction of the exchange interaction, which leads to a weak repulsion between the Co impurity and the skyrmion. This simple explanation is consistent with the rests obtained by the DFT calculations in Fig.~\ref{fig3}.

Finally we provide an estimate of $J$ in Eq. \eqref{eqH} based on the DFT results. The calculated skyrmion energy relative to the ferromagnetic state is about $\sim$ 0.84 meV /f.u., therefore the skyrmion energy in the $8\times8$ mesh is about $\sim$ 54 meV  ($\sim$ 540 K). The skyrmion self-energy in Eq. \eqref{eqH}  is of the order of $J$. The N\'{e}el temperature $T_N$ is of the order of $J N_c$ with $N_c$ the coordination number of spins. For MnSi according to experiments, $T_N\approx 45$ K \cite{Yu2010a} and the estimated $J$ is one of order of magnitude smaller than that of DFT calculations. There are two reasons: firstly the DFT calculations overestimate the magnetic moment by a factor of $2.5$ and thus overestimate the energy by a factor of $6.25$; secondly, the thermal and quantum fluctuations that are not accounted for in the DFT calculations can reduce the mean-field $T_N$ by one order of magnitude. After taking these two effects into account, the estimated $J$ by the DFT calculations and experiments can be comparable within the same order of magnitude.

\section{Summary}
We have studied the effect of an atomic impurity on the skyrmion in MnSi by performing the density functional theory calculations. We have demonstrated that the interactions between the defects and skyrmion can be tuned by substitution of different elements. For Mn substituted by Zn or Ir and Si substituted by Pb, the interaction is attractive, indicating a pinning of the skyrmion. While for Mn substituted by Co, the interaction between the defect and the skyrmion is weakly repulsive. We have also computed the local density of state and magnetic moments to understand the impacts of impurities on the magnetic and electronic properties. Based on the density functional theory results, we have provided a qualitative understanding using a phenomenological model. Due to the limitation of the density functional theory for a large spin texture, the calculations are subjected to the finite size finite effect. Nevertheless we believe that the qualitative features presented here should still hold in the large system size.

%\begin{figure}[tb]
%\includegraphics[width=1.0\linewidth]{fig1.eps}
%\caption{(color online)
%The $T$-dependent FS evolution in the DFT+DMFT calculation.
%}
%\label{fig1}
%\end{figure}

%two column  figure
%\begin{figure*}[tb]
%\includegraphics[width=1.0\linewidth]{fig2.eps}
%\caption{(color online)
%The $T$-dependent dHvA frequencies ($\mathrm{F}$) and cyclotron effective masses ($m^*$).
%}
%\label{fig2}
%\end{figure*}

\begin{acknowledgments}
We thank Yuan-Yen Tai for helpful discussions. This work was carried out under the auspices of the National Nuclear Security Administration of the US DOE at LANL under Contract No. DE-AC52-06NA25396 and was supported by the LANL LDRD-DR Program.
\end{acknowledgments}

\bibliography{reference}

\begin{thebibliography}{29}%
\makeatletter
\providecommand \@ifxundefined [1]{%
 \@ifx{#1\undefined}
}%
\providecommand \@ifnum [1]{%
 \ifnum #1\expandafter \@firstoftwo
 \else \expandafter \@secondoftwo
 \fi
}%
\providecommand \@ifx [1]{%
 \ifx #1\expandafter \@firstoftwo
 \else \expandafter \@secondoftwo
 \fi
}%
\providecommand \natexlab [1]{#1}%
\providecommand \enquote  [1]{``#1''}%
\providecommand \bibnamefont  [1]{#1}%
\providecommand \bibfnamefont [1]{#1}%
\providecommand \citenamefont [1]{#1}%
\providecommand \href@noop [0]{\@secondoftwo}%
\providecommand \href [0]{\begingroup \@sanitize@url \@href}%
\providecommand \@href[1]{\@@startlink{#1}\@@href}%
\providecommand \@@href[1]{\endgroup#1\@@endlink}%
\providecommand \@sanitize@url [0]{\catcode `\\12\catcode `\$12\catcode
  `\&12\catcode `\#12\catcode `\^12\catcode `\_12\catcode `\%12\relax}%
\providecommand \@@startlink[1]{}%
\providecommand \@@endlink[0]{}%
\providecommand \url  [0]{\begingroup\@sanitize@url \@url }%
\providecommand \@url [1]{\endgroup\@href {#1}{\urlprefix }}%
\providecommand \urlprefix  [0]{URL }%
\providecommand \Eprint [0]{\href }%
\providecommand \doibase [0]{http://dx.doi.org/}%
\providecommand \selectlanguage [0]{\@gobble}%
\providecommand \bibinfo  [0]{\@secondoftwo}%
\providecommand \bibfield  [0]{\@secondoftwo}%
\providecommand \translation [1]{[#1]}%
\providecommand \BibitemOpen [0]{}%
\providecommand \bibitemStop [0]{}%
\providecommand \bibitemNoStop [0]{.\EOS\space}%
\providecommand \EOS [0]{\spacefactor3000\relax}%
\providecommand \BibitemShut  [1]{\csname bibitem#1\endcsname}%
\let\auto@bib@innerbib\@empty
%</preamble>
\bibitem [{\citenamefont {Bogdanov}\ and\ \citenamefont
  {Yablonskii}(1989)}]{Bogdanov89}%
  \BibitemOpen
  \bibfield  {author} {\bibinfo {author} {\bibfnamefont {A.~N.}\ \bibnamefont
  {Bogdanov}}\ and\ \bibinfo {author} {\bibfnamefont {D.~A.}\ \bibnamefont
  {Yablonskii}},\ }\href@noop {} {\bibfield  {journal} {\bibinfo  {journal}
  {Sov. Phys. JETP}\ }\textbf {\bibinfo {volume} {68}},\ \bibinfo {pages} {101}
  (\bibinfo {year} {1989})}\BibitemShut {NoStop}%
\bibitem [{\citenamefont {M\"{u}hlbauer}\ \emph {et~al.}(2009)\citenamefont
  {M\"{u}hlbauer}, \citenamefont {Binz}, \citenamefont {Jonietz}, \citenamefont
  {Pfleiderer}, \citenamefont {Rosch}, \citenamefont {Neubauer}, \citenamefont
  {Georgii},\ and\ \citenamefont {B\"{o}ni}}]{Muhlbauer2009}%
  \BibitemOpen
  \bibfield  {author} {\bibinfo {author} {\bibfnamefont {S.}~\bibnamefont
  {M\"{u}hlbauer}}, \bibinfo {author} {\bibfnamefont {B.}~\bibnamefont {Binz}},
  \bibinfo {author} {\bibfnamefont {F.}~\bibnamefont {Jonietz}}, \bibinfo
  {author} {\bibfnamefont {C.}~\bibnamefont {Pfleiderer}}, \bibinfo {author}
  {\bibfnamefont {A.}~\bibnamefont {Rosch}}, \bibinfo {author} {\bibfnamefont
  {A.}~\bibnamefont {Neubauer}}, \bibinfo {author} {\bibfnamefont
  {R.}~\bibnamefont {Georgii}}, \ and\ \bibinfo {author} {\bibfnamefont
  {P.}~\bibnamefont {B\"{o}ni}},\ }\href {\doibase 10.1126/science.1166767}
  {\bibfield  {journal} {\bibinfo  {journal} {Science}\ }\textbf {\bibinfo
  {volume} {323}},\ \bibinfo {pages} {915} (\bibinfo {year}
  {2009})}\BibitemShut {NoStop}%
\bibitem [{\citenamefont {Yu}\ \emph {et~al.}(2010)\citenamefont {Yu},
  \citenamefont {Onose}, \citenamefont {Kanazawa}, \citenamefont {Park},
  \citenamefont {Han}, \citenamefont {Matsui}, \citenamefont {Nagaosa},\ and\
  \citenamefont {Tokura}}]{Yu2010a}%
  \BibitemOpen
  \bibfield  {author} {\bibinfo {author} {\bibfnamefont {X.~Z.}\ \bibnamefont
  {Yu}}, \bibinfo {author} {\bibfnamefont {Y.}~\bibnamefont {Onose}}, \bibinfo
  {author} {\bibfnamefont {N.}~\bibnamefont {Kanazawa}}, \bibinfo {author}
  {\bibfnamefont {J.~H.}\ \bibnamefont {Park}}, \bibinfo {author}
  {\bibfnamefont {J.~H.}\ \bibnamefont {Han}}, \bibinfo {author} {\bibfnamefont
  {Y.}~\bibnamefont {Matsui}}, \bibinfo {author} {\bibfnamefont
  {N.}~\bibnamefont {Nagaosa}}, \ and\ \bibinfo {author} {\bibfnamefont
  {Y.}~\bibnamefont {Tokura}},\ }\href {\doibase 10.1038/nature09124}
  {\bibfield  {journal} {\bibinfo  {journal} {Nature}\ }\textbf {\bibinfo
  {volume} {465}},\ \bibinfo {pages} {901} (\bibinfo {year}
  {2010})}\BibitemShut {NoStop}%
\bibitem [{\citenamefont {Yu}\ \emph {et~al.}(2011)\citenamefont {Yu},
  \citenamefont {Kanazawa}, \citenamefont {Onose}, \citenamefont {Kimoto},
  \citenamefont {Zhang}, \citenamefont {Ishiwata}, \citenamefont {Matsui},\
  and\ \citenamefont {Tokura}}]{Yu2011}%
  \BibitemOpen
  \bibfield  {author} {\bibinfo {author} {\bibfnamefont {X.~Z.}\ \bibnamefont
  {Yu}}, \bibinfo {author} {\bibfnamefont {N.}~\bibnamefont {Kanazawa}},
  \bibinfo {author} {\bibfnamefont {Y.}~\bibnamefont {Onose}}, \bibinfo
  {author} {\bibfnamefont {K.}~\bibnamefont {Kimoto}}, \bibinfo {author}
  {\bibfnamefont {W.~Z.}\ \bibnamefont {Zhang}}, \bibinfo {author}
  {\bibfnamefont {S.}~\bibnamefont {Ishiwata}}, \bibinfo {author}
  {\bibfnamefont {Y.}~\bibnamefont {Matsui}}, \ and\ \bibinfo {author}
  {\bibfnamefont {Y.}~\bibnamefont {Tokura}},\ }\href {\doibase
  10.1038/nmat2916} {\bibfield  {journal} {\bibinfo  {journal} {Nat. Mater}\
  }\textbf {\bibinfo {volume} {10}},\ \bibinfo {pages} {106} (\bibinfo {year}
  {2011})}\BibitemShut {NoStop}%
\bibitem [{\citenamefont {Adams}\ \emph {et~al.}(2012)\citenamefont {Adams},
  \citenamefont {Chacon}, \citenamefont {Wagner}, \citenamefont {Bauer},
  \citenamefont {Brandl}, \citenamefont {Pedersen}, \citenamefont {Berger},
  \citenamefont {Lemmens},\ and\ \citenamefont {Pfleiderer}}]{Adams2012}%
  \BibitemOpen
  \bibfield  {author} {\bibinfo {author} {\bibfnamefont {T.}~\bibnamefont
  {Adams}}, \bibinfo {author} {\bibfnamefont {A.}~\bibnamefont {Chacon}},
  \bibinfo {author} {\bibfnamefont {M.}~\bibnamefont {Wagner}}, \bibinfo
  {author} {\bibfnamefont {A.}~\bibnamefont {Bauer}}, \bibinfo {author}
  {\bibfnamefont {G.}~\bibnamefont {Brandl}}, \bibinfo {author} {\bibfnamefont
  {B.}~\bibnamefont {Pedersen}}, \bibinfo {author} {\bibfnamefont
  {H.}~\bibnamefont {Berger}}, \bibinfo {author} {\bibfnamefont
  {P.}~\bibnamefont {Lemmens}}, \ and\ \bibinfo {author} {\bibfnamefont
  {C.}~\bibnamefont {Pfleiderer}},\ }\href {\doibase
  10.1103/PhysRevLett.108.237204} {\bibfield  {journal} {\bibinfo  {journal}
  {Phys. Rev. Lett.}\ }\textbf {\bibinfo {volume} {108}},\ \bibinfo {pages}
  {237204} (\bibinfo {year} {2012})}\BibitemShut {NoStop}%
\bibitem [{\citenamefont {Seki}\ \emph {et~al.}(2012)\citenamefont {Seki},
  \citenamefont {Yu}, \citenamefont {Ishiwata},\ and\ \citenamefont
  {Tokura}}]{Seki2012}%
  \BibitemOpen
  \bibfield  {author} {\bibinfo {author} {\bibfnamefont {S.}~\bibnamefont
  {Seki}}, \bibinfo {author} {\bibfnamefont {X.~Z.}\ \bibnamefont {Yu}},
  \bibinfo {author} {\bibfnamefont {S.}~\bibnamefont {Ishiwata}}, \ and\
  \bibinfo {author} {\bibfnamefont {Y.}~\bibnamefont {Tokura}},\ }\href
  {\doibase 10.1126/science.1214143} {\bibfield  {journal} {\bibinfo  {journal}
  {Science}\ }\textbf {\bibinfo {volume} {336}},\ \bibinfo {pages} {198}
  (\bibinfo {year} {2012})}\BibitemShut {NoStop}%
\bibitem [{\citenamefont {Dzyaloshinsky}(1958)}]{Dzyaloshinsky1958}%
  \BibitemOpen
  \bibfield  {author} {\bibinfo {author} {\bibfnamefont {I.}~\bibnamefont
  {Dzyaloshinsky}},\ }\href {\doibase 10.1016/0022-3697(58)90076-3} {\bibfield
  {journal} {\bibinfo  {journal} {J. Phys. Chem. Solids}\ }\textbf {\bibinfo
  {volume} {4}},\ \bibinfo {pages} {241} (\bibinfo {year} {1958})}\BibitemShut
  {NoStop}%
\bibitem [{\citenamefont {Moriya}(1960{\natexlab{a}})}]{Moriya60}%
  \BibitemOpen
  \bibfield  {author} {\bibinfo {author} {\bibfnamefont {T.}~\bibnamefont
  {Moriya}},\ }\href {\doibase 10.1103/PhysRev.120.91} {\bibfield  {journal}
  {\bibinfo  {journal} {Phys. Rev.}\ }\textbf {\bibinfo {volume} {120}},\
  \bibinfo {pages} {91} (\bibinfo {year} {1960}{\natexlab{a}})}\BibitemShut
  {NoStop}%
\bibitem [{\citenamefont {Moriya}(1960{\natexlab{b}})}]{Moriya60b}%
  \BibitemOpen
  \bibfield  {author} {\bibinfo {author} {\bibfnamefont {T.}~\bibnamefont
  {Moriya}},\ }\href {\doibase 10.1103/PhysRevLett.4.228} {\bibfield  {journal}
  {\bibinfo  {journal} {Phys. Rev. Lett.}\ }\textbf {\bibinfo {volume} {4}},\
  \bibinfo {pages} {228} (\bibinfo {year} {1960}{\natexlab{b}})}\BibitemShut
  {NoStop}%
\bibitem [{\citenamefont {Jonietz}\ \emph {et~al.}(2010)\citenamefont
  {Jonietz}, \citenamefont {M\"uhlbauer}, \citenamefont {Pfleiderer},
  \citenamefont {Neubauer}, \citenamefont {M\"unzer}, \citenamefont {Bauer},
  \citenamefont {Adams}, \citenamefont {Georgii}, \citenamefont {B\"oni},
  \citenamefont {Duine}, \citenamefont {Everschor}, \citenamefont {Garst},\
  and\ \citenamefont {Rosch}}]{Jonietz2010}%
  \BibitemOpen
  \bibfield  {author} {\bibinfo {author} {\bibfnamefont {F.}~\bibnamefont
  {Jonietz}}, \bibinfo {author} {\bibfnamefont {S.}~\bibnamefont
  {M\"uhlbauer}}, \bibinfo {author} {\bibfnamefont {C.}~\bibnamefont
  {Pfleiderer}}, \bibinfo {author} {\bibfnamefont {A.}~\bibnamefont
  {Neubauer}}, \bibinfo {author} {\bibfnamefont {W.}~\bibnamefont {M\"unzer}},
  \bibinfo {author} {\bibfnamefont {A.}~\bibnamefont {Bauer}}, \bibinfo
  {author} {\bibfnamefont {T.}~\bibnamefont {Adams}}, \bibinfo {author}
  {\bibfnamefont {R.}~\bibnamefont {Georgii}}, \bibinfo {author} {\bibfnamefont
  {P.}~\bibnamefont {B\"oni}}, \bibinfo {author} {\bibfnamefont {R.~A.}\
  \bibnamefont {Duine}}, \bibinfo {author} {\bibfnamefont {K.}~\bibnamefont
  {Everschor}}, \bibinfo {author} {\bibfnamefont {M.}~\bibnamefont {Garst}}, \
  and\ \bibinfo {author} {\bibfnamefont {A.}~\bibnamefont {Rosch}},\ }\href
  {\doibase 10.1126/science.1195709} {\bibfield  {journal} {\bibinfo  {journal}
  {Science}\ }\textbf {\bibinfo {volume} {330}},\ \bibinfo {pages} {1648}
  (\bibinfo {year} {2010})}\BibitemShut {NoStop}%
\bibitem [{\citenamefont {Yu}\ \emph {et~al.}(2012)\citenamefont {Yu},
  \citenamefont {Kanazawa}, \citenamefont {Zhang}, \citenamefont {Nagai},
  \citenamefont {Hara}, \citenamefont {Kimoto}, \citenamefont {Matsui},
  \citenamefont {Onose},\ and\ \citenamefont {Tokura}}]{Yu2012}%
  \BibitemOpen
  \bibfield  {author} {\bibinfo {author} {\bibfnamefont {X.~Z.}\ \bibnamefont
  {Yu}}, \bibinfo {author} {\bibfnamefont {N.}~\bibnamefont {Kanazawa}},
  \bibinfo {author} {\bibfnamefont {W.~Z.}\ \bibnamefont {Zhang}}, \bibinfo
  {author} {\bibfnamefont {T.}~\bibnamefont {Nagai}}, \bibinfo {author}
  {\bibfnamefont {T.}~\bibnamefont {Hara}}, \bibinfo {author} {\bibfnamefont
  {K.}~\bibnamefont {Kimoto}}, \bibinfo {author} {\bibfnamefont
  {Y.}~\bibnamefont {Matsui}}, \bibinfo {author} {\bibfnamefont
  {Y.}~\bibnamefont {Onose}}, \ and\ \bibinfo {author} {\bibfnamefont
  {Y.}~\bibnamefont {Tokura}},\ }\href {\doibase 10.1038/ncomms1990} {\bibfield
   {journal} {\bibinfo  {journal} {Nat. Commun.}\ }\textbf {\bibinfo {volume}
  {3}},\ \bibinfo {pages} {988} (\bibinfo {year} {2012})}\BibitemShut {NoStop}%
\bibitem [{\citenamefont {Schulz}\ \emph {et~al.}(2012)\citenamefont {Schulz},
  \citenamefont {Ritz}, \citenamefont {Bauer}, \citenamefont {Halder},
  \citenamefont {Wagner}, \citenamefont {Franz}, \citenamefont {Pfleiderer},
  \citenamefont {Everschor}, \citenamefont {Garst},\ and\ \citenamefont
  {Rosch}}]{Schulz2012}%
  \BibitemOpen
  \bibfield  {author} {\bibinfo {author} {\bibfnamefont {T.}~\bibnamefont
  {Schulz}}, \bibinfo {author} {\bibfnamefont {R.}~\bibnamefont {Ritz}},
  \bibinfo {author} {\bibfnamefont {A.}~\bibnamefont {Bauer}}, \bibinfo
  {author} {\bibfnamefont {M.}~\bibnamefont {Halder}}, \bibinfo {author}
  {\bibfnamefont {M.}~\bibnamefont {Wagner}}, \bibinfo {author} {\bibfnamefont
  {C.}~\bibnamefont {Franz}}, \bibinfo {author} {\bibfnamefont
  {C.}~\bibnamefont {Pfleiderer}}, \bibinfo {author} {\bibfnamefont
  {K.}~\bibnamefont {Everschor}}, \bibinfo {author} {\bibfnamefont
  {M.}~\bibnamefont {Garst}}, \ and\ \bibinfo {author} {\bibfnamefont
  {A.}~\bibnamefont {Rosch}},\ }\href {\doibase 10.1038/nphys2231} {\bibfield
  {journal} {\bibinfo  {journal} {Nat. Phys.}\ }\textbf {\bibinfo {volume}
  {8}},\ \bibinfo {pages} {301} (\bibinfo {year} {2012})}\BibitemShut {NoStop}%
\bibitem [{\citenamefont {White}\ \emph {et~al.}(2012)\citenamefont {White},
  \citenamefont {Levatic}, \citenamefont {Omrani}, \citenamefont {Egetenmeyer},
  \citenamefont {Prsa}, \citenamefont {Zivkovic}, \citenamefont {Gavilano},
  \citenamefont {Kohlbrecher}, \citenamefont {Bartkowiak}, \citenamefont
  {Berger},\ and\ \citenamefont {Ronnow}}]{White2012}%
  \BibitemOpen
  \bibfield  {author} {\bibinfo {author} {\bibfnamefont {J.~S.}\ \bibnamefont
  {White}}, \bibinfo {author} {\bibfnamefont {I.}~\bibnamefont {Levatic}},
  \bibinfo {author} {\bibfnamefont {A.~A.}\ \bibnamefont {Omrani}}, \bibinfo
  {author} {\bibfnamefont {N.}~\bibnamefont {Egetenmeyer}}, \bibinfo {author}
  {\bibfnamefont {K.}~\bibnamefont {Prsa}}, \bibinfo {author} {\bibfnamefont
  {I.}~\bibnamefont {Zivkovic}}, \bibinfo {author} {\bibfnamefont {J.~L.}\
  \bibnamefont {Gavilano}}, \bibinfo {author} {\bibfnamefont {J.}~\bibnamefont
  {Kohlbrecher}}, \bibinfo {author} {\bibfnamefont {M.}~\bibnamefont
  {Bartkowiak}}, \bibinfo {author} {\bibfnamefont {H.}~\bibnamefont {Berger}},
  \ and\ \bibinfo {author} {\bibfnamefont {H.~M.}\ \bibnamefont {Ronnow}},\
  }\href {\doibase 10.1088/0953-8984/24/43/432201} {\bibfield  {journal}
  {\bibinfo  {journal} {J. Phys.: Condens. Matter}\ }\textbf {\bibinfo {volume}
  {24}},\ \bibinfo {pages} {432201} (\bibinfo {year} {2012})}\BibitemShut
  {NoStop}%
\bibitem [{\citenamefont {White}\ \emph {et~al.}(2014)\citenamefont {White},
  \citenamefont {Pr\ifmmode~\check{s}\else \v{s}\fi{}a}, \citenamefont {Huang},
  \citenamefont {Omrani}, \citenamefont {\ifmmode \check{Z}\else
  \v{Z}\fi{}ivkovi\ifmmode~\acute{c}\else \'{c}\fi{}}, \citenamefont
  {Bartkowiak}, \citenamefont {Berger}, \citenamefont {Magrez}, \citenamefont
  {Gavilano}, \citenamefont {Nagy}, \citenamefont {Zang},\ and\ \citenamefont
  {R\o{}nnow}}]{PhysRevLett.113.107203}%
  \BibitemOpen
  \bibfield  {author} {\bibinfo {author} {\bibfnamefont {J.~S.}\ \bibnamefont
  {White}}, \bibinfo {author} {\bibfnamefont {K.}~\bibnamefont
  {Pr\ifmmode~\check{s}\else \v{s}\fi{}a}}, \bibinfo {author} {\bibfnamefont
  {P.}~\bibnamefont {Huang}}, \bibinfo {author} {\bibfnamefont {A.~A.}\
  \bibnamefont {Omrani}}, \bibinfo {author} {\bibfnamefont {I.}~\bibnamefont
  {\ifmmode \check{Z}\else \v{Z}\fi{}ivkovi\ifmmode~\acute{c}\else
  \'{c}\fi{}}}, \bibinfo {author} {\bibfnamefont {M.}~\bibnamefont
  {Bartkowiak}}, \bibinfo {author} {\bibfnamefont {H.}~\bibnamefont {Berger}},
  \bibinfo {author} {\bibfnamefont {A.}~\bibnamefont {Magrez}}, \bibinfo
  {author} {\bibfnamefont {J.~L.}\ \bibnamefont {Gavilano}}, \bibinfo {author}
  {\bibfnamefont {G.}~\bibnamefont {Nagy}}, \bibinfo {author} {\bibfnamefont
  {J.}~\bibnamefont {Zang}}, \ and\ \bibinfo {author} {\bibfnamefont {H.~M.}\
  \bibnamefont {R\o{}nnow}},\ }\href {\doibase 10.1103/PhysRevLett.113.107203}
  {\bibfield  {journal} {\bibinfo  {journal} {Phys. Rev. Lett.}\ }\textbf
  {\bibinfo {volume} {113}},\ \bibinfo {pages} {107203} (\bibinfo {year}
  {2014})}\BibitemShut {NoStop}%
\bibitem [{\citenamefont {Shibata}\ \emph {et~al.}(2015)\citenamefont
  {Shibata}, \citenamefont {Iwasaki}, \citenamefont {Kanazawa}, \citenamefont
  {Aizawa}, \citenamefont {Tanigaki}, \citenamefont {Shirai}, \citenamefont
  {Nakajima}, \citenamefont {Kubota}, \citenamefont {Kawasaki}, \citenamefont
  {Park}, \citenamefont {Shindo}, \citenamefont {Nagaosa},\ and\ \citenamefont
  {Tokura}}]{Shibata2015}%
  \BibitemOpen
  \bibfield  {author} {\bibinfo {author} {\bibfnamefont {K.}~\bibnamefont
  {Shibata}}, \bibinfo {author} {\bibfnamefont {J.}~\bibnamefont {Iwasaki}},
  \bibinfo {author} {\bibfnamefont {N.}~\bibnamefont {Kanazawa}}, \bibinfo
  {author} {\bibfnamefont {S.}~\bibnamefont {Aizawa}}, \bibinfo {author}
  {\bibfnamefont {T.}~\bibnamefont {Tanigaki}}, \bibinfo {author}
  {\bibfnamefont {M.}~\bibnamefont {Shirai}}, \bibinfo {author} {\bibfnamefont
  {T.}~\bibnamefont {Nakajima}}, \bibinfo {author} {\bibfnamefont
  {M.}~\bibnamefont {Kubota}}, \bibinfo {author} {\bibfnamefont
  {M.}~\bibnamefont {Kawasaki}}, \bibinfo {author} {\bibfnamefont {H.~S.}\
  \bibnamefont {Park}}, \bibinfo {author} {\bibfnamefont {D.}~\bibnamefont
  {Shindo}}, \bibinfo {author} {\bibfnamefont {N.}~\bibnamefont {Nagaosa}}, \
  and\ \bibinfo {author} {\bibfnamefont {Y.}~\bibnamefont {Tokura}},\ }\href
  {\doibase 10.1038/nnano.2015.113} {\bibfield  {journal} {\bibinfo  {journal}
  {Nat. Nanotechnol.}\ }\textbf {\bibinfo {volume} {10}},\ \bibinfo {pages}
  {589} (\bibinfo {year} {2015})}\BibitemShut {NoStop}%
\bibitem [{\citenamefont {Fert}\ \emph {et~al.}(2013)\citenamefont {Fert},
  \citenamefont {Cros},\ and\ \citenamefont {Sampaio}}]{Fert2013}%
  \BibitemOpen
  \bibfield  {author} {\bibinfo {author} {\bibfnamefont {A.}~\bibnamefont
  {Fert}}, \bibinfo {author} {\bibfnamefont {V.}~\bibnamefont {Cros}}, \ and\
  \bibinfo {author} {\bibfnamefont {J.}~\bibnamefont {Sampaio}},\ }\href
  {\doibase 10.1038/nnano.2013.29} {\bibfield  {journal} {\bibinfo  {journal}
  {Nat. Nanotechnol.}\ }\textbf {\bibinfo {volume} {8}},\ \bibinfo {pages}
  {152} (\bibinfo {year} {2013})}\BibitemShut {NoStop}%
\bibitem [{\citenamefont {Zhou}\ and\ \citenamefont
  {Ezawa}(2014)}]{zhou_reversible_2014}%
  \BibitemOpen
  \bibfield  {author} {\bibinfo {author} {\bibfnamefont {Y.}~\bibnamefont
  {Zhou}}\ and\ \bibinfo {author} {\bibfnamefont {M.}~\bibnamefont {Ezawa}},\
  }\href {\doibase 10.1038/ncomms5652} {\bibfield  {journal} {\bibinfo
  {journal} {Nat. Commun.}\ }\textbf {\bibinfo {volume} {5}},\ \bibinfo {pages}
  {4652} (\bibinfo {year} {2014})}\BibitemShut {NoStop}%
\bibitem [{\citenamefont {Sampaio}\ \emph {et~al.}(2013)\citenamefont
  {Sampaio}, \citenamefont {Cros}, \citenamefont {Rohart}, \citenamefont
  {Thiaville},\ and\ \citenamefont {Fert}}]{sampaio_nucleation_2013}%
  \BibitemOpen
  \bibfield  {author} {\bibinfo {author} {\bibfnamefont {J.}~\bibnamefont
  {Sampaio}}, \bibinfo {author} {\bibfnamefont {V.}~\bibnamefont {Cros}},
  \bibinfo {author} {\bibfnamefont {S.}~\bibnamefont {Rohart}}, \bibinfo
  {author} {\bibfnamefont {A.}~\bibnamefont {Thiaville}}, \ and\ \bibinfo
  {author} {\bibfnamefont {A.}~\bibnamefont {Fert}},\ }\href {\doibase
  10.1038/nnano.2013.210} {\bibfield  {journal} {\bibinfo  {journal} {Nat.
  Nanotechnol.}\ }\textbf {\bibinfo {volume} {8}},\ \bibinfo {pages} {839}
  (\bibinfo {year} {2013})}\BibitemShut {NoStop}%
\bibitem [{\citenamefont {Blatter}\ \emph {et~al.}(1994)\citenamefont
  {Blatter}, \citenamefont {Feigel'man}, \citenamefont {Geshkenbein},
  \citenamefont {Larkin},\ and\ \citenamefont {Vinokur}}]{Blatter94}%
  \BibitemOpen
  \bibfield  {author} {\bibinfo {author} {\bibfnamefont {G.}~\bibnamefont
  {Blatter}}, \bibinfo {author} {\bibfnamefont {M.~V.}\ \bibnamefont
  {Feigel'man}}, \bibinfo {author} {\bibfnamefont {V.~B.}\ \bibnamefont
  {Geshkenbein}}, \bibinfo {author} {\bibfnamefont {A.~I.}\ \bibnamefont
  {Larkin}}, \ and\ \bibinfo {author} {\bibfnamefont {V.~M.}\ \bibnamefont
  {Vinokur}},\ }\href {\doibase 10.1103/RevModPhys.66.1125} {\bibfield
  {journal} {\bibinfo  {journal} {Rev. Mod. Phys.}\ }\textbf {\bibinfo {volume}
  {66}},\ \bibinfo {pages} {1125} (\bibinfo {year} {1994})}\BibitemShut
  {NoStop}%
\bibitem [{\citenamefont {Lin}\ \emph {et~al.}(2013)\citenamefont {Lin},
  \citenamefont {Reichhardt}, \citenamefont {Batista},\ and\ \citenamefont
  {Saxena}}]{szlin13skyrmion2}%
  \BibitemOpen
  \bibfield  {author} {\bibinfo {author} {\bibfnamefont {S.-Z.}\ \bibnamefont
  {Lin}}, \bibinfo {author} {\bibfnamefont {C.}~\bibnamefont {Reichhardt}},
  \bibinfo {author} {\bibfnamefont {C.~D.}\ \bibnamefont {Batista}}, \ and\
  \bibinfo {author} {\bibfnamefont {A.}~\bibnamefont {Saxena}},\ }\href
  {\doibase 10.1103/PhysRevB.87.214419} {\bibfield  {journal} {\bibinfo
  {journal} {Phys. Rev. B}\ }\textbf {\bibinfo {volume} {87}},\ \bibinfo
  {pages} {214419} (\bibinfo {year} {2013})}\BibitemShut {NoStop}%
\bibitem [{\citenamefont {Everschor}\ \emph {et~al.}(2012)\citenamefont
  {Everschor}, \citenamefont {Garst}, \citenamefont {Binz}, \citenamefont
  {Jonietz}, \citenamefont {M\"uhlbauer}, \citenamefont {Pfleiderer},\ and\
  \citenamefont {Rosch}}]{Everschor12}%
  \BibitemOpen
  \bibfield  {author} {\bibinfo {author} {\bibfnamefont {K.}~\bibnamefont
  {Everschor}}, \bibinfo {author} {\bibfnamefont {M.}~\bibnamefont {Garst}},
  \bibinfo {author} {\bibfnamefont {B.}~\bibnamefont {Binz}}, \bibinfo {author}
  {\bibfnamefont {F.}~\bibnamefont {Jonietz}}, \bibinfo {author} {\bibfnamefont
  {S.}~\bibnamefont {M\"uhlbauer}}, \bibinfo {author} {\bibfnamefont
  {C.}~\bibnamefont {Pfleiderer}}, \ and\ \bibinfo {author} {\bibfnamefont
  {A.}~\bibnamefont {Rosch}},\ }\href {\doibase 10.1103/PhysRevB.86.054432}
  {\bibfield  {journal} {\bibinfo  {journal} {Phys. Rev. B}\ }\textbf {\bibinfo
  {volume} {86}},\ \bibinfo {pages} {054432} (\bibinfo {year}
  {2012})}\BibitemShut {NoStop}%
\bibitem [{\citenamefont {Iwasaki}\ \emph {et~al.}(2013)\citenamefont
  {Iwasaki}, \citenamefont {Mochizuki},\ and\ \citenamefont
  {Nagaosa}}]{Iwasaki2013}%
  \BibitemOpen
  \bibfield  {author} {\bibinfo {author} {\bibfnamefont {J.}~\bibnamefont
  {Iwasaki}}, \bibinfo {author} {\bibfnamefont {M.}~\bibnamefont {Mochizuki}},
  \ and\ \bibinfo {author} {\bibfnamefont {N.}~\bibnamefont {Nagaosa}},\ }\href
  {\doibase 10.1038/ncomms2442} {\bibfield  {journal} {\bibinfo  {journal}
  {Nat. Commun.}\ }\textbf {\bibinfo {volume} {4}},\ \bibinfo {pages} {1463}
  (\bibinfo {year} {2013})}\BibitemShut {NoStop}%
\bibitem [{\citenamefont {Kresse}\ and\ \citenamefont {Joubert}(1999)}]{paw}%
  \BibitemOpen
  \bibfield  {author} {\bibinfo {author} {\bibfnamefont {G.}~\bibnamefont
  {Kresse}}\ and\ \bibinfo {author} {\bibfnamefont {D.}~\bibnamefont
  {Joubert}},\ }\href {\doibase 10.1103/PhysRevB.59.1758} {\bibfield  {journal}
  {\bibinfo  {journal} {Phys. Rev. B}\ }\textbf {\bibinfo {volume} {59}},\
  \bibinfo {pages} {1758} (\bibinfo {year} {1999})}\BibitemShut {NoStop}%
\bibitem [{\citenamefont {Kresse}\ and\ \citenamefont
  {Furthm\"uller}(1996)}]{vasp}%
  \BibitemOpen
  \bibfield  {author} {\bibinfo {author} {\bibfnamefont {G.}~\bibnamefont
  {Kresse}}\ and\ \bibinfo {author} {\bibfnamefont {J.}~\bibnamefont
  {Furthm\"uller}},\ }\href {\doibase 10.1103/PhysRevB.54.11169} {\bibfield
  {journal} {\bibinfo  {journal} {Phys. Rev. B}\ }\textbf {\bibinfo {volume}
  {54}},\ \bibinfo {pages} {11169} (\bibinfo {year} {1996})}\BibitemShut
  {NoStop}%
\bibitem [{\citenamefont {Ishikawa}\ \emph {et~al.}(1976)\citenamefont
  {Ishikawa}, \citenamefont {Tajima}, \citenamefont {Bloch},\ and\
  \citenamefont {Roth}}]{Ishikawa1976}%
  \BibitemOpen
  \bibfield  {author} {\bibinfo {author} {\bibfnamefont {T.}~\bibnamefont
  {Ishikawa}}, \bibinfo {author} {\bibfnamefont {K.}~\bibnamefont {Tajima}},
  \bibinfo {author} {\bibfnamefont {P.}~\bibnamefont {Bloch}}, \ and\ \bibinfo
  {author} {\bibfnamefont {M.}~\bibnamefont {Roth}},\ }\href {\doibase
  10.1016/0038-1098(76)90057-0} {\bibfield  {journal} {\bibinfo  {journal}
  {Solid State Commun.}\ }\textbf {\bibinfo {volume} {19}},\ \bibinfo {pages}
  {525} (\bibinfo {year} {1976})}\BibitemShut {NoStop}%
\bibitem [{\citenamefont {Jeong}\ and\ \citenamefont
  {Pickett}(2004)}]{MnSi-DFT-04}%
  \BibitemOpen
  \bibfield  {author} {\bibinfo {author} {\bibfnamefont {T.}~\bibnamefont
  {Jeong}}\ and\ \bibinfo {author} {\bibfnamefont {W.~E.}\ \bibnamefont
  {Pickett}},\ }\href {\doibase 10.1103/PhysRevB.70.075114} {\bibfield
  {journal} {\bibinfo  {journal} {Phys. Rev. B}\ }\textbf {\bibinfo {volume}
  {70}},\ \bibinfo {pages} {075114} (\bibinfo {year} {2004})}\BibitemShut
  {NoStop}%
\bibitem [{\citenamefont {Okamura}\ \emph {et~al.}(2013)\citenamefont
  {Okamura}, \citenamefont {Kagawa}, \citenamefont {Mochizuki}, \citenamefont
  {Kubota}, \citenamefont {Seki}, \citenamefont {Ishiwata}, \citenamefont
  {Kawasaki}, \citenamefont {Onose},\ and\ \citenamefont
  {Tokura}}]{okamura_microwave_2013}%
  \BibitemOpen
  \bibfield  {author} {\bibinfo {author} {\bibfnamefont {Y.}~\bibnamefont
  {Okamura}}, \bibinfo {author} {\bibfnamefont {F.}~\bibnamefont {Kagawa}},
  \bibinfo {author} {\bibfnamefont {M.}~\bibnamefont {Mochizuki}}, \bibinfo
  {author} {\bibfnamefont {M.}~\bibnamefont {Kubota}}, \bibinfo {author}
  {\bibfnamefont {S.}~\bibnamefont {Seki}}, \bibinfo {author} {\bibfnamefont
  {S.}~\bibnamefont {Ishiwata}}, \bibinfo {author} {\bibfnamefont
  {M.}~\bibnamefont {Kawasaki}}, \bibinfo {author} {\bibfnamefont
  {Y.}~\bibnamefont {Onose}}, \ and\ \bibinfo {author} {\bibfnamefont
  {Y.}~\bibnamefont {Tokura}},\ }\href {\doibase 10.1038/ncomms3391} {\bibfield
   {journal} {\bibinfo  {journal} {Nat. Commun.}\ }\textbf {\bibinfo {volume}
  {4}},\ \bibinfo {pages} {2391} (\bibinfo {year} {2013})}\BibitemShut
  {NoStop}%
\bibitem [{\citenamefont {Schwarze}\ \emph {et~al.}(2015)\citenamefont
  {Schwarze}, \citenamefont {Waizner}, \citenamefont {Garst}, \citenamefont
  {Bauer}, \citenamefont {Stasinopoulos}, \citenamefont {Berger}, \citenamefont
  {Pfleiderer},\ and\ \citenamefont {Grundler}}]{schwarze_universal_2015}%
  \BibitemOpen
  \bibfield  {author} {\bibinfo {author} {\bibfnamefont {T.}~\bibnamefont
  {Schwarze}}, \bibinfo {author} {\bibfnamefont {J.}~\bibnamefont {Waizner}},
  \bibinfo {author} {\bibfnamefont {M.}~\bibnamefont {Garst}}, \bibinfo
  {author} {\bibfnamefont {A.}~\bibnamefont {Bauer}}, \bibinfo {author}
  {\bibfnamefont {I.}~\bibnamefont {Stasinopoulos}}, \bibinfo {author}
  {\bibfnamefont {H.}~\bibnamefont {Berger}}, \bibinfo {author} {\bibfnamefont
  {C.}~\bibnamefont {Pfleiderer}}, \ and\ \bibinfo {author} {\bibfnamefont
  {D.}~\bibnamefont {Grundler}},\ }\href {\doibase 10.1038/nmat4223} {\bibfield
   {journal} {\bibinfo  {journal} {Nat. Mater.}\ }\textbf {\bibinfo {volume}
  {14}},\ \bibinfo {pages} {478} (\bibinfo {year} {2015})}\BibitemShut
  {NoStop}%
\bibitem [{\citenamefont {Bogdanov}\ and\ \citenamefont
  {Hubert}(1994)}]{Bogdanov94}%
  \BibitemOpen
  \bibfield  {author} {\bibinfo {author} {\bibfnamefont {A.}~\bibnamefont
  {Bogdanov}}\ and\ \bibinfo {author} {\bibfnamefont {A.}~\bibnamefont
  {Hubert}},\ }\href {\doibase http://dx.doi.org/10.1016/0304-8853(94)90046-9}
  {\bibfield  {journal} {\bibinfo  {journal} {J. Magn. Magn. Mater.}\ }\textbf
  {\bibinfo {volume} {138}},\ \bibinfo {pages} {255 } (\bibinfo {year}
  {1994})}\BibitemShut {NoStop}%
\end{thebibliography}%

\end{document}